Short Paper

# A Unique One-Time Password Table Sequence Pattern Authentication: Application to Bicol University Union of Federated Faculty Association, Inc. (BUUFFAI) eVoting System

Benedicto B. Balilo Jr.
Technological Institute of the Philippines
Quezon City, Philippines
benedicto.balilojr@gmail.com

Bobby D. Gerardo
West Visayas State University
Iloilo City, Philippines
bobby.gerardo@gmail.com

Ruji P. Medina
Dean, Technological Institute of the Philippines Graduate Programs
Quezon City, Philippines
ruji_p_medina@yahoo.com

Yungcheol Byun
Faculty of Telecommunications and Computer Engineering
Cheju National University, Jeju, Korea
ycb@jejunu.ac.kr

**Abstract**

*Purpose* – Electronic Voting System (EVS) is a type of voting program that deals primarily with the selection, the casting of votes with embedded security mechanism that detects errors, and the tamper-proof election of results done through the use of an electronic system. It can include optical scan, specialized voting kiosks and Internet voting approach. Most organizations have difficulties when it comes to voting and the Bicol University Union of Federated Faculty Association Incorporated (BUUFFAI) is not an exception. Some of the problems involved include convenience, cost, geographical location of the polling precinct, and voting turnouts.

*Method* – This study extends the scope of the current BUUFFAI eVoting system to address such issues and to eliminate inconvenience both to the faculty voters and the facilitators. This voting scheme used an algorithmic OTP scheme based on table sequence pattern schedule that randomly generates an XY coordinate unique to voters that will be sent to voter registered email address.

*Results* – This study addressed the security requirements and maintained election procedures with confidentiality, integrity and availability.

# INTRODUCTION

Electronic voting systems (EVS) are widely used in elections. The EVS (also known as eVoting System) encompasses different types of voting that used electronic means to count votes. There are two types of e-voting system: the paper voting and the remote voting. Paper voting includes the current voting system and postal voting (absentee voting). Electronic Voting technology can include optical scan voting systems and specialized voting kiosks (Lee, Lee, Won, & Kim, 2010). When voters cast their votes, the latter are recorded to the chip and the removable memory card can serve as storage or temporary backup. EVS can also include transmission of ballots and votes via telephones, private computer network or the internet. The voting stations may or may not be interconnected and may operate as a single unit as a complete polling station. The system tracks the number of votes and continuously displays the number of votes cast on a counter system. In terms of security, it can be controlled by the polling authorities or operated independently (Kotob et al., 2004).

Riera and Brown (2003) believed that the lack of transparency in electronic voting systems can be overcome with extent technology, physical, and procedure security measures. They believed that such measures provides clarity to the process and avoid the need to rely on complex and/or network systems and/or proprietary closed systems, and minimize the number of components that must be trusted to software generating the encrypted ballot and the software opening of digital boxes. These would avoid problems interposed by the computer systems and technical personnel between the voter and the electoral board. Nonetheless, doubts still remain that some criteria must be completed and satisfied. Some of the basic criteria include confidentiality, integrity, availability, reliability and assurance for computer systems (Neumann, 1993).

Olaniyi, Arulogun, and Omidiora (2013) developed an eVoting system designed to improve the authentication and integrity of evoting system multifactor authentication and cryptographic hash function methods. It gathered the two key security issues to secure e-voting systems; namely, threat of erring voter's authentication and integrity of vote transmitted over insecure wireless medium.

The BUUFFAI is an association of teaching and non-teaching staff of the Bicol University. Election is conducted every four (4) years and the elected president will sit as faculty representative in the Board of Regents (BOR). The association constitution and by-laws provide the procedures of the election process including the organization of election committee, campaign activities and others. Faculty members shall be active members (regular faculty member) of the university and must be in a good standing status (paid annual union dues). Currently, the BUUFFAI has been implementing the network-based voting system. It was implemented as a solution to the shortcomings of the manual election procedure.

The Bicol University has campuses, namely, Polangui, Tabaco, Gubat, East and West campuses. Faculty members from these campuses need to come to the designated election precincts to vote with access to computers and technical support. Each faculty member upon registration is given a reference code which serves as entry to the system. The system together with the reference code verifies and authenticates the user and prompted to the list of candidates. The automation follows the casting and printing of results. In its implementation, the opportunity of online system was recognized that will allow members to cast their votes without compromising the integrity of the results and security of the system.

The purpose of this study is to introduce the opportunity offered by an online eVoting system that will take advantage of the use of the technology resources of the university, increase faculty voter's turnout, introduce convenience to union members, and improve the generation of reference codes during election with integrity and secure authentication procedure. The introduction of OTP will increase the level of security and efficiency of the system without additional cost and effort to the organization. This may serve as opportunity for the integration of the existing LAN-Based Election system and the consideration of the Board. The study sought to answer the following questions: 1) What information requirements are needed in the proposed system? 2) What are the features of the proposed system? and 3) What security mechanism shall be adopted in the proposed system?

# RELATED WORKS

Electronic voting process is considered as one of the easiest and fastest means to cast votes and automatically generate election reports. It offers the advantages of (1) eliminating the possibility of invalid and doubtful votes which, in many cases, are the root causes of controversies and election petitions, (2) making the process of counting of votes faster than the conventional system, (3) reducing the quantity of paper used, and (4) reduce cost of printing (Kumar & Walia, 2011).

In today's trend, the internet has become a widely popular tool for eVoting activities. It facilitates the process similar to an online survey which simplifies the process of organizing elections and make it convenient for voters to vote remotely from their home computer while taking into consideration security, anonymity at the same time provide auditioning capabilities. This will save time, effort and paper work (Bila et al., 2005). However, preference concern shall be provided over its vulnerabilities, threats, and lapses.

In order to address these concerns, Kim, Lee, and Lee (2015) studied the vulnerabilities that overcome eVoting system. They concluded that a double audit should be done at the levels of the producer and electoral district with other forces included such as observers, national and municipal monitoring as to the accuracy and independence of eVoting machine functionality. Human based approach, knowledge diversity, and securing the software are parts of the audit to detect possible faulty system. Platform compatibility, deployment, user interface design, logical voter's status, backup, security and presentation of results (Bringula et al., 2012) are among the lapses that should be given considerations.

In terms of security, Kohno, Stubblefield, Rubin, and Wallach (2004) identified several problems including unauthorized privilege escalation, incorrect use of cryptography, vulnerabilities to network threats, and poor software development processes which are serious threats to both insider and outsider attacks. To ensure security on voting process, Al-Anie, Alia, & Hnaif (2011) concluded that the establishment of a protocol based on RSA public-key encryption cryptosystem increased the level of security and efficiency of the system that allows voters to vote using the basic requirements only such as personal computer (PC), internet connection, voting website and standard mobile phone without any extra cost and effort. Different researchers implemented different methods to address security concerns. Maeung (2006) presented 3 related crypto-graphical theories; namely, homomorphic encryption, efficient honest-verifier zero-knowledge, and threshold decryption cryptosystem to compute and verify both voters and authorities. That is, to allow convenient and confident voting while maintaining the accuracy of election results.

Olaniyi, Folorunso, Ahmed, & Joseph (2016) applied the unimodal fingerprint biometrics and advanced encryption standard based on Wavelet crypto-watermarking approach. It solved the possibility of blundering voter's authentication, integrity and confidentiality of vote stored in the server. Furthermore, the results revealed that after the evaluation using anti-watermarking detectors, the system served as a platform for the possible delivery of credible e-election.

Pandit, Bhawar, and Desai (2014) developed a system that is capable of denying access to any unauthorized voter and of preventing multiple votes from the same voter by employing steganography in the voting process. They concluded that the integrity of the voting process lies from the minute the voter casts the vote until the cast vote is registered. The number of important functional and non-functional requirements gives a voter the option of viewing candidate profile and provides easy use of the system.

Al-Ameen and Talab (2013) studied and analyzed the technical attributes of a good eVoting system and the reason for each attribute with respect to the voting process which leads to the introduction of new concept in internet voting system. They concluded that eVoting becomes the quickest, cheapest, and the most efficient way to administer election and count vote as it only consists of simple a process or procedure. However, the system is not free from threats that may hinder the system from functioning correctly. They recommended that before applying the system, there should be a thorough consideration and fulfilment of the requirements and application of a multi-testing process.

Nowadays, mobile phones for electronic voting are widely-used during elections. It can be used as a tool to send notifications or messages directly to the user. Also, it can be a tool to authenticate a user from accessing confidential and sensitive information. Akhare et al. (2016) described mobile phones as efficient tools for voting. Mobile phones can be used for registration process irrespective of user location. For example, in Thailand, the Thai Election Commission sent messages to 25 million users reminding them to vote (Stein, 2011).

Many vendors have developed commercial voting systems with different levels of functionalities like AccuVote TS, eSlate300, STAR-Vote (Bell et al., 2013), iVotronic and many others. However, these commercial applications have limitations and functional properties that may not fit organizational requirements like student council setup, technical resources and technology acceptance.

The proposed study aims to introduce an eVoting system based on OTP table sequence pattern schedule authentication scheme for voter's identification that will address issues encountered during BUUFFAI election and present a new election system concept not only to speed up the electoral procedures but also to address convenience, resource maximization, minimizing labor, and reduction of cost., and make it possible for faculty to conveniently vote despite different geographical campus locations.

# THE PROPOSED METHOD

The proposed design covers a synchronized OTP, security mechanism based on email authentication model, audit trail logs and enhanced voting procedures (see Figure 1). Access to the system requires identification and verification of voter information. A faculty member through the registration module presents the full details including the college or department and status to participate in the voting process.

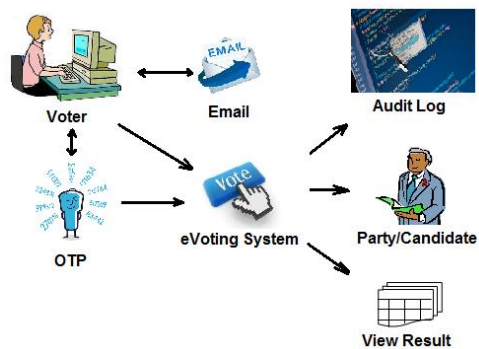

*Figure 1*. Diagram of the Proposed eVoting System

In the proposed system, only qualified voter can access the eVoting system through identification and authentication process. The username and password will serve as the first level of authentication process covered in the registration module (see Figure 2). In voting, voters can request a Reference OTP code as access to display the voting process window.

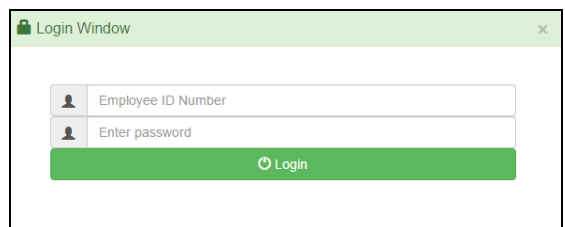

*Figure 2*. Login Window of the Proposed System

A faculty member can request reference code to avail of the OTP code. As a requirement, the voter shall input the correct Employee ID Number for verification (see Figure 3). The system will validate if the faculty member is an active union member and is qualified (in good standing) to vote. Once verified, the system generates OTP codes and these will be sent to the email address of the voter. If the inputted OTP codes match the generated codes, the voter can use of the voting window. To secure and protect the integrity of the results, an audit trail log is incorporated. The log will be the trigger event and basis that will state that the voter performed access to the system.

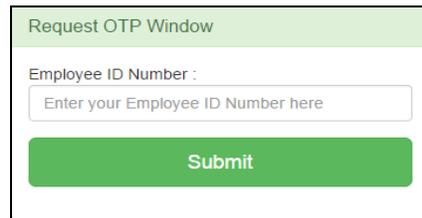

*Figure 3.* Request OTP Window of the Proposed System

The introduction and integration of the algorithm into the existing system shall increase individual participation, ensure convenience, and save time and effort as faculty members can vote without leaving their station/campus and cast votes at their convenient time.

The Rational Unified Process (RUP) methodology as guide in the development of the proposed eVoting system was used in the study. It is expected that with the use of the set of building blocks and content elements including those that are to be produced and the necessary skills required giving a step-by-step explanation of specific development goals, the RUP will position the development process to identify its milestone, artefacts, and cut-off unnecessary results. The RUP deals with a lifecycle that ends with a milestone. Each cycle in the RUP methodology is broken down into sequence of four phases, such as, Inception, Elaboration, Construction and Transition.

In order to gather appropriate information, and determine the information needed for the study, the researchers used interviews, document analysis and observation for data gathering.

# RESULTS AND DISCUSSION

In the above simulation, the system was tested in Intel Corei5-4460 processor based running at 3.210GHz with 2GB of RAM.

*A. Information Requirements Needed in the Proposed System*

Information is fundamental in every undertaking or development system process. This information served as the basis for analysis requirements leading to the development of the design features of the system. The information requirements include the 2014-2016 list of candidates, current system process adopted by the association and member records. These are the primary information requirements to determine the projected features and functionalities of the proposed system.

The information requirements provide clear identification of the problems, opportunities, and project objectives of the proposed system. The report generation module extends to support the existing system over graphical summary of cast votes and turnovered results.

*B. Features of the Proposed System*

- *User Registration and Login Module*

Registration is the process of recording users' information. In the user registration module, the employee ID, password, lastname, firstname, college, position, contact number, email address and reference code (generated values) are the attributes needed in order to be registered in the system. The college and position are dynamically set for easy access and to avoid typographical errors. The contact number can be any local number capable of receiving messages. The process will be verified and approved by an Administrator (an individual designated by the association through a board resolution). In the login process, the username and password are the primary components needed to manipulate the system.

- *Home Page of the Proposed System*

The BUUFFAI GUI has diffirent modules of the proposed system (see Figure 4). These include voter registration, candidate information, login, and tutorial module. Each module has its level of functionality that will address the voters' request. The voter registration module allows faculty members to store their respective personal details and institution information. These pieces of information will be the basis to determine the current standing of the faculty member in the BUUFFAI.

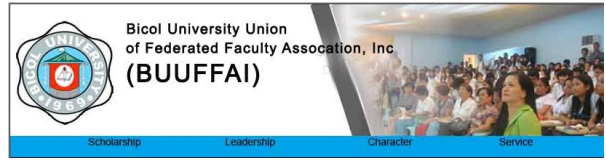

*Figure 4*. BUUFFAI Home Page

The faculty member will be guided through the candidate information module to provide information as regards to candidate qualifications, programs and platforms.

- *Voter OTP Authentication*

The generated OTP code sent through faculty registered email address would be the official login entry into the system. This OTP code is unique to every faculty member who requested an online voting procedure. The generation of OTP is one way of addressing the security requirements of the system and maintain election procedures with confidentiality, integrity and availability

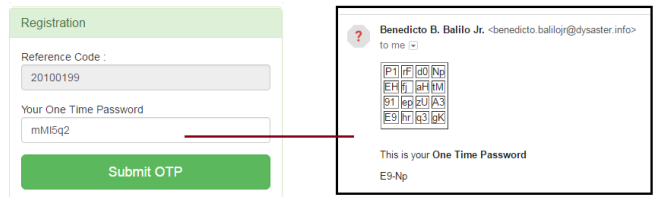

*Figure 5*. OTP Voter Verification sent via Email

Figure 5 shows the OTP code sent to voter registered email address. The purpose of sending the randomly-generated XY coordinate is to free the user from brute force, dictionary, and insider attack because the voter has to do physical matching of the XY coordinates and input the full details of the code into the OTP input box (labelled "Your One Time Password") for server authentication. The server authenticates whether the inputted OTP matches the generated OTP. Once authenticated, the system displays the candidate information window and the voter could select from the candidate list and automatically send the selection by clicking the button labelled "Submit OTP".

Figure 6 indicates that the user has received an SMS message with the label 'INFO' showing the details of the OTP codes. The SMS message captures the XY axis and automatically determines the values; thus, displays the six (6) OTP values.

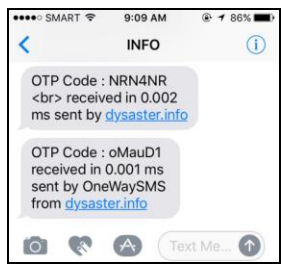

*Figure 6*. Sample SMS messages received by User

The generation of OTP codes sent to faculty member email address and registered contact number (SMS) are the unique features of the system providing security advantage over the existing system (LAN-Based).

- *Election Page*

This shows the actual election process of the proposed system. In this page, voters can select their preferred candidates from the list and view the initial summary results of votes per position/candidate (see Figure 7). The election module presents the actual voting procedures. Through the setting module, faculty members can add, edit, update and view information.

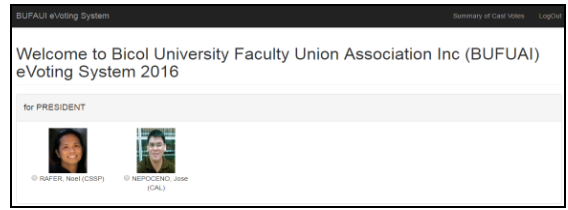

*Figure 7.* Candidate Selection Window

The candidates' picture and position together with the selection button is displayed to provide the voter a graphical interface of the candidates. To submit the entries, voters shall click the button labelled "Cast Votes". This process is a one-time selection; therefore, voters can submit and vote once. This allows the system to check a flag marking to the voter database for audit trail purposes, security to the system, and integrity to the results of the election.

C. *Proposed OTP authentication scheme*

OTP is a combination of strings of characters, numbers and symbols (see Formula 1). The algorithm used the concept of random number generation, attribute-based and string manipulation technique. The initial seed is represented by letter *g* to be the *string of characters + numbers + date + timestamp.* Using the formula below, the initial seed captures the current OTP and is integrated as part of the next OTP to be generated. The letter b represents the OTP to be generated.

$$b=g,\ b^1=(b(g+otp^1),\ b^2=(b^1(g+otp^2))..\ b^{n+1}=(b^{n+1}(g+otp^n)) \qquad Formula\ 1$$

Figure 8 shows the generated two (2) pair values extracted from the initial seed values *abcefghijklmnopqrstuvwxyzABCDEFGHIJKLMNOPQRSTUVWXYZApr04Mon1020171491809436*.

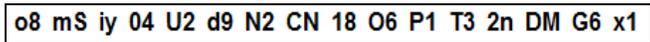

*Figure 8.* Generated OTP values

These values will be mapped out into the table together with the randomly-generated XY values, given the formula *OTP = xy( PassCode, Combi (aA, Num),2)*. The interesting part of the research is the application and combination of these values producing a two (2) pair code that will be mapped out in a 4x4 table using a sequence schedule algorithm. Figure 9 shows the algorithm and how the OTP codes will be generated.

The initial seed will be dynamically-generated from the combination of parameters (Line 3) together with the current date and time (Line 5). The core of the algorithm is the generation of randomize codes which will be stored in an array as temporary storage (Line 10). The mapping of values returns the row and column element (Line18).

| Line | Pseudo code |
|------|-------------|
| 1 | Function generateOTP |
| 2 | Begin |
| 3 | Set predetermined characters and numbers |
| 4 | Set two-dimensional array |
| 5 | Get current date and time |
| 6 | Merge = Array1, Array2 |
| 7 | While i < 15 Do |
| 8 | Randomize the array |
| 9 | Extract two values |
| 10 | Store in array |
| 11 | End While |
| 12 | End |
| 13 | |
| 14 | Function generateXY |
| 15 | Begin |
| 16 | Get row element |
| 17 | Get column element |
| 18 | return row element , column element |
| 19 | End |

*Figure 9*. OTP Pseudo Code of the Proposed System

The 4x4 sequence pattern schedule is a table of rows and columns with a combination of two (2) pairs of alphanumeric codes. It is important not to repeat the OTP codes; array randomization scheme was embedded to the algorithm itself (shown in Figure 10).

| 04 | o8 | O6 | U2 |
| d9 | T3 | 18 | 2n |
| DM | CN | mS | x1 |
| iy | N2 | G6 | P1 |

*Figure 10.* Sample Generated OTP in 4x4 matrix

The placing of the codes follows the sequential pattern schedule using left-right (LF) posting. The reading from the array used the First-in Last-out (FILO) queue method. This method will be free from brute force and dictionary attack as the applied algorithmic pattern uses the combination of randomized characters and values with randomly generated One Time Password codes.

### *D. Audit Log*

From the voters' request to the submission of the candidate item lists, the proposed system secure a log trail of each voter credentials. The reference code generated by the system that was sent to the voter registered email address triggers the event that an activity for electronic voting procedure is about to commence. The proposed system tag or marked the voters' credential as potential user for such activity; thus, the OTP is ready waiting for activation.

Once the reference code and button event signal the action, the OTP codes will be generated and will open the access to deliver the OTP codes via email system. The system will wait for the exit trigger event that will allow the process to collect, record, and store detailed information into the database.

## CONCLUSIONS AND RECOMMENDATIONS

The application of Electronic voting system has been used by/in many organizations, associations or even in the highest hierarchy of the government. The information requirements can lead and guide system developers to identify what possible response should be carried out in times of election given the experiences, feedback, election results, and post election assessment.

The proposed system together with its features will serve as opportunity for the association especially the officers to consider, evaluate, and link the integration to the existing system. Decrease in faculty voter turnout can be observed as the system shall address issues concerning geographical location and take trip to designated election venues. This will bring opportunities to the association especially to members for the developed system to be accessed with minimum technology requirement.

Because of increase advantage, voters should be literate in the use of the current technology. The proposed electronic voting system has embedded security mechanism that is free from attacks and threats. Moreover, the EVS would make it possible to provide voter's identity and to transmit election results electronically. Furthermore, the system can speed up the election process, lessen cost, provide security, maximize organization's resources and introduce convenience. The introduction of OTP email-based authentication scheme was adequate enough to safeguard and protect, and not hamper the reliability of the information in the system.

As a recommendation, the system should be tested using black box and white box testing including system testing and should pass through an appropriate evaluation of concern before its final integration to the existing system. Likewise, trainings should be given to future administrators and wide dissemination must be encouraged.

## ACKNOWLEDGEMENTS

Sincere gratitude is given to Doctor Sony Valdez and Prof. Romina Villamor of Aquinas University, Legazpi City.